# Absence of a COVID-induced academic drop in high-school physics learning

Eric W. Burkholder, *Department of Physics Auburn University, Auburn AL USA 36849*
Carl E. Wieman, *Department of Physics, Stanford University, Stanford CA USA 94305*

**Abstract:** At the start of the COVID-19 pandemic, the majority of secondary instruction in the United States transitioned to an online environment. In many parts of the country, online schooling continued for upwards of two years. Many experts have hypothesized an "academic slide" – a reduction in student learning – following this period of online instruction. We investigated the change in student preparation for introductory college physics in incoming Stanford university students between the Fall term of 2019 to the Fall term of 2021. We did this by looking at the performance on a validated physics diagnostic exam that all Stanford students intending to take a physics course took before enrolling in an introductory physics course. We found no statistically or educationally significant change in scores. Despite many anecdotal faculty reports, at least for this population the level of student preparation in physics and related math appears to be unchanged.

**Introduction:**

There have been many reports in the popular media in which university faculty claim that the incoming college students are much less prepared now than they were a few years ago due to the disruptions of COVID-19 on education [1]. We have also heard similar statements from faculty at our institutions. While there are clearly many serious negative consequences from COVID on students of all levels, this is the type of situation where perceptions and evidence are strongly shaped by confirmation bias[2]. This motivated us to try and test these claims. A recent study showed an average reduction in ACT scores in the U.S., but it is unclear what this indicates about college preparation. The reported change was small, and this reduction many have been due to many schools waiving standardized test requirements during the pandemic[3]. We realized that we had data that would allow us to objectively measure the difference in the directly relevant incoming preparation in physics for Stanford undergraduates who completed high school pre- and post-COVID.

The Stanford undergraduate population is not representative of the population of US college students. The institution is highly selective, with the fraction of applicants who are admitted trending down slowly over the years but remaining essentially constant before and after COVID. While this population represents a small and high performing fraction of the high school population, it is the same fraction before and after COVID, and physics preparation counts very little in the admissions process. This is evident in the very wide distribution of scores shown below on the physics diagnostic exam. Compared to the US university undergraduate population at large, the Stanford population is moderately diverse. So, if there was a reduction in physics and related math learning that was similar across the entire population of graduating high school students in the US, we would expect to see the effect on these Stanford students. That outcome would be consistent with the ACT's reports that drops in performance occur across all groups of students[3]. However, if there was a detrimental impact only on the low but not the high performing students across the US, we would not see it in this study, because our sample only includes the latter.

**Methods:**

Students who have not yet taken a physics course at Stanford University are required to take an empirically validated physics diagnostic exam[4] to provide them with a recommendation as to which of the different introductory physics course they should take. They typically take this exam the summer before they start college. The diagnostic exam covers a range of topics in physics and mathematics, including vector operations, basic integration, Taylor series approximations, kinematics, static equilibrium, Newton's laws, conservation of energy, angular momentum, and rotational motion. It is strongly predictive of the student scores in the courses. Students were not awarded credit for completing the exam, but they were required to complete it before they could enroll in an introductory physics course.

We examined scores collected in the summer of 2019 (prior to the COVID-19 pandemic) and the summer of 2021 (after most students had experienced two years of remote instruction). We removed those few students from our sample (<1%) who spent less than 10 minutes completing the exam. The median time spent on the assessment was 38 minutes, indicating that most students were making a serious effort. This left 881 students in the 2019 sample and 968 students in the 2021 sample. (The 2021 sample is likely larger because an unusually large number of students admitted for 2020 chose to defer starting college until 2021.) We calculated the mean and standard deviation separately for students who intended to enroll in algebra-based and calculus-based physics, as this provides an objective way to look for possible changes in the top and bottom of the overall distribution separately. Students with better preparation in physics and math look to take the calculus-based course. For each of the two populations (calculus and non-calculus based) we tested for differences in the averages in 2019 compared with 2021 using a t-test. We also calculated Cohen's d between the two years and used histograms to qualitatively investigate the differences in the respective distributions.

The Stanford student population is 28% White, 25% Asian, 18% Hispanic or Latino, 7% Black or African American, 11% Multiracial, 11% International students, and <3% Native American or Pacific Islander. Approximately 33% of students are first generation/low-income students – students with no parent that received a four-year college degree. The interquartile range of admitted students' ACT scores is 32-35.

**Results:**

The distribution of scores for algebra-based physics is plotted in Fig. 1. In 2019, the average score was 17.3 out of 36 (standard deviation = 6.63 points; $N = 214$). In 2021, the average score was 17.1 points (s. d. = 6.89 points; $N = 212$). The difference between these two means as a fraction of the standard deviation is $d = -0.027$ and is not statistically significant ($p = 0.78$).

The distribution of scores for calculus-based physics is plotted in Fig. 2. In 2019, the average score was 24.1 out of 36 (standard deviation = 7.44 points; $N = 597$). In 2021, the average score was 23.6 points (s. d. = 7.31 points; $N = 757$). The difference between these two means is $d = -0.061$ and is not statistically significant ($p = 0.27$).

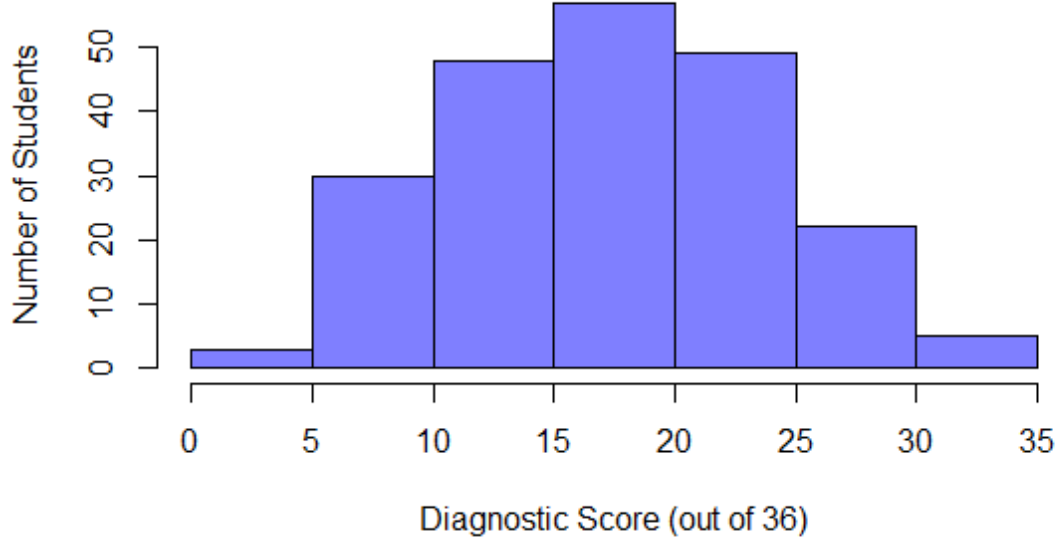

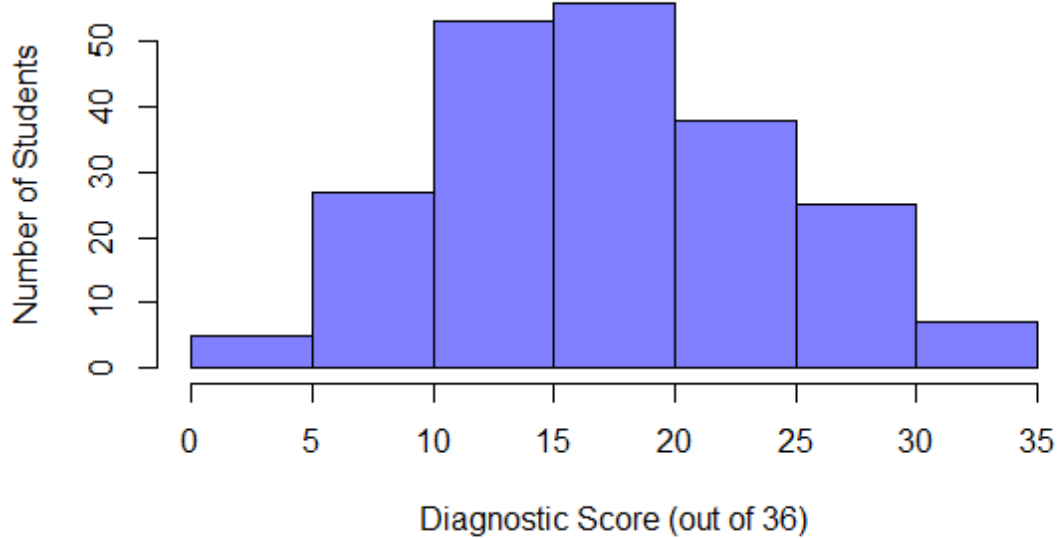

*Figure 1: Score distribution in 2019 (top) and 2021 (bottom) for students intending to enroll in algebra-based physics.*

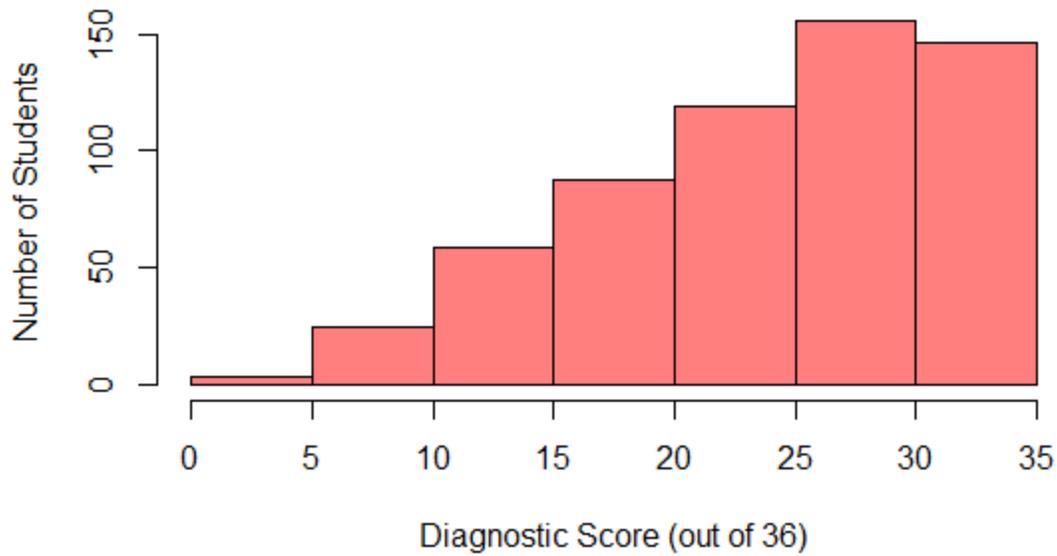

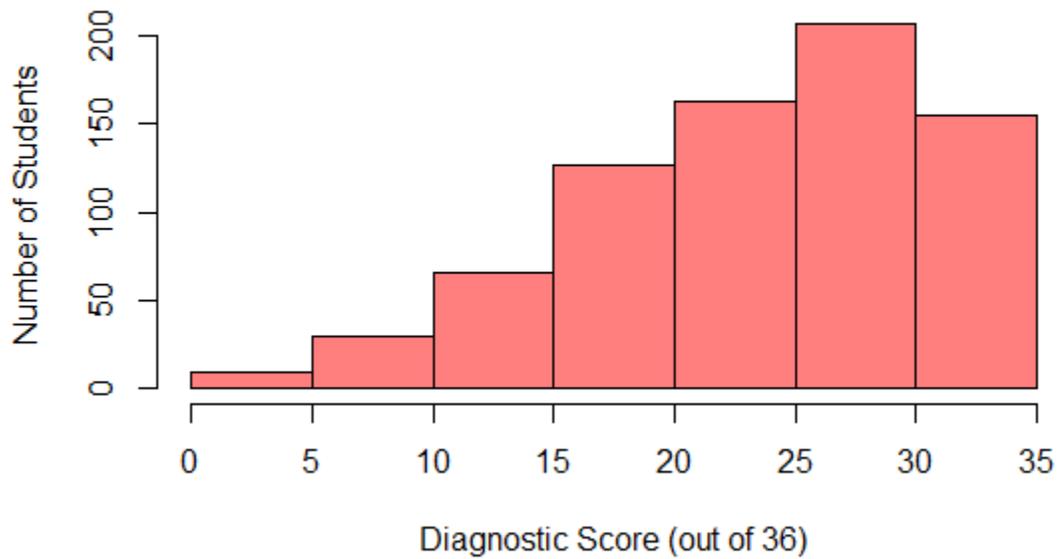

*Figure 2: Score distribution in 2019 (top) and 2021 (bottom) for students intending to enroll in calculus-based physics.*

**Discussion:**

We observe essentially zero change in scores between 2019 and 2021, the differences in the means are a tiny fraction of the standard deviations and completely consistent with random fluctuations. The standard deviations are also nearly identical between 2019 and 2021.

These results show that, at least in the domain of physics, this group of students showed no detrimental impact on learning due to COVID. In contrast to predictions[5] and many faculty perceptions[1] this suggests that the negative effects of remote instruction on student *learning* are small, at least for this population.

This is an academically high-achieving population overall and may not be representative of the broader U. S. population. Many of these students come from well-resourced high schools and households and may have had the resources to reduce any negative impact of remote learning. However, we know from earlier studies that a substantial fraction of these students are from backgrounds and poorly resourced high schools in which they are not receiving good instruction in physics[6]. However, we do not observe any shift in the score distributions at the lower end, where students from under-resourced schools and households typically score on this assessment. These are the students that are expected to suffer the most educationally from the impacts of covid.

We emphasize that we are only measuring understanding of math and physics, and not any of the other variables that are known to relate to overall student performance and mental health, such as anxiety, motivation, self-efficacy, etc.

**Conclusions:**

The impact of the pandemic on Stanford students' academic preparation has been smaller than expected or perceived. This suggests that university instructors in introductory physics may not need to reduce the level of their courses as many have suggested.

We do not wish to suggest that the overall impact of the pandemic on students is minor. Many students will have lost parents and relatives, suffered intense social isolation, and feared for their own well-being for an extended period of time. All of these factors are expected to have significant tolls on students' mental health and academic success. There are numerous indications that student mental health has suffered significantly at both our institutions and many others. Our message is only that some anticipated negative impacts of the pandemic may not be as large as expected and more careful analysis than faculty anecdotal impressions is needed.

---

[1] e.g, "My College Students Are Not OK", Jonathan Malesic, NY Times, May 13, 2022. A google search turns up many more.
[2] Nickerson, Raymond S. (1998), "Confirmation bias: A ubiquitous phenomenon in many guises", *Review of General Psychology*, **2** (2): 175–220
[3] Hayes, S. (2021). Learning opportunities: Understanding scores from ACT's assessment suite during the COVID-19 pandemic. *ACT*.
[4] Burkholder, E. W., K. D. Wang, and C. E. Wieman (2021). Validated diagnostic test for introductory physics course placement. *Physical Review Physics Education Research.*